# Experimental observation of the elusive double-peak structure in R-dependent strong-field ionization rate of $H_2^+$

Han Xu[1,*], Feng He[2,#], D. Kielpinski[1,3], R.T. Sang[1,3], I.V. Litvinyuk[1,†]

[1] *Centre for Quantum Dynamics and Australian Attosecond Science Facility, Griffith University, Nathan, QLD 4111, Australia*
[2] *Key Laboratory for Laser Plasmas (Ministry of Education), and Department of Physics and Astronomy, IFSA Collaborative Innovation Center, SJTU, Shanghai 200240, People's Republic of China*
[3] *ARC Centre of Excellence for Coherent X-Ray Science, Griffith University, Nathan, QLD 4111, Australia*
[*] *hanxu1981@gmail.com*
[#] *fhe@sjtu.edu.cn*
[†] *i.litvinyuk@griffith.edu.au*

**When a diatomic molecule is ionized by an intense laser field, the ionization rate depends very strongly on the inter-nuclear separation. That dependence exhibits a pronounced maximum at the inter-nuclear separation known as the "critical distance". This phenomenon was first demonstrated theoretically in $H_2^+$ and became known as "charge-resonance enhanced ionization" (CREI, in reference to a proposed physical mechanism) or simply "enhanced ionization"(EI). All theoretical models of this phenomenon predict a double-peak structure in the R-dependent ionization rate of $H_2^+$. However, such double-peak structure has never been observed experimentally. It was even suggested that it is impossible to observe due to fast motion of the nuclear wavepackets. Here we report a few-cycle pump-probe experiment which clearly resolves that elusive double-peak structure. In the experiment, an expanding $H_2^+$ ion produced by an intense pump pulse is probed by a much weaker probe pulse. The predicted double-peak structure is clearly seen in delay-dependent kinetic energy spectra of protons when pump and probe pulses are polarized parallel to each other. No structure is seen when the probe is polarized perpendicular to the pump.**

The ionization of atoms and molecules by strong laser fields is fundamentally important for understanding laser-matter interactions in general. The ionization of molecules is generally much more complex than that of atoms, since ionization processes depend sensitively on nuclear coordinates, such as molecular alignment and inter-nuclear distances. It has been found, both experimentally [1-6] and theoretically [7-11] that the ionization rate of a diatomic molecule can be dramatically enhanced (by orders of magnitude) when its inter-nuclear separation reaches so-called critical distance. This phenomenon appears to be a universal feature of molecular ionization, and it was recently demonstrated in tri- [12] and poly-atomic [13] molecules.

The fundamental physical mechanism of EI is well understood for $H_2^+$ [9-11] (see figure 1(b)). When $H_2^+$ with its double-well potential is exposed to a static electric field the electron localized in the upper potential well will experience potential barrier suppressed by the Coulomb field of the second proton and it can much more easily tunnel out than an atomic electron. However, for this to happen in an oscillating electric field of a laser pulse the electron first has to be localized in the upper potential well. Therefore, both electron and nuclear dynamics are playing a role in the interaction: the inter-nuclear distance needs to be large enough for substantial electron density to remain trapped in the upper potential well yet small enough for significant suppression of the internal barrier. While simple physical reasoning readily accounts for existence of a critical distance, quite curiously, all theoretical calculations of various degrees of sophistication predict not one but two adjacent maxima for ionization rate in the critical region of R. The origin of this double-peak structure remains obscure and its very existence until now had no experimental confirmation. Obviously, the very concept of R-dependent ionization rate is only meaningful in the context of the fixed-nuclei approximation. So naturally, failure to observe the double-peak structure predicted by the fixed-nuclei theoretical models was interpreted as a failure of the fixed-nuclei approximation itself for fast-moving and spatially-extended nuclear wavepackets in $H_2^+$. Therefore, our observation of the predicted double-peak structure supports the validity of the fixed-nuclei approximation in this specific context.

To measure the R-dependent ionization rate in $H_2^+$ one must generate the target molecular ion with different (and well controlled) distributions of inter-nuclear separations. That could only be achieved in a pump-probe experiment, where the pump would ionize neutral hydrogen and generate nuclear dynamics in the resulting molecular ion, which could then be further ionized by a delayed probe pulse. Since this nuclear dynamics is very fast (it takes about 15 fs for the wavepacket to reach the outer turning point) both pulses have to be very short – no more than 2 cycles in duration at 800 nm wavelength (6 fs). Several such experiments have been performed [4,5] and failed to register the double-peak structure. It has been suggested that the "wash out" effect due to fast nuclear motion [5], and the fast depletion of $H_2^+$ at the first peak [14], make it hard, if not impossible, to observe the double-peak structure. It should be pointed out that these experiments were performed with pump and probe pulses having the same or very similar peak intensities, sufficiently high to ionize neutral hydrogen molecules with significant probability. However, it is predicted that $H_2^+$ at the critical distance is much easier to ionize than $H_2$ at its equilibrium R. So one might expect that probe pulses of excessive intensity would deplete $H_2^+$ well before reaching their peak power, thus making the observation of the minimum between the two peaks impossible. Keeping that in mind, we performed a pump-probe experiment using ultra-short (6 fs) pulses, with probe pulse being an order of magnitude less intense than the pump pulse (pump - $6\times10^{14}$ W/cm$^2$, probe - $6\times10^{13}$ W/cm$^2$). Under such conditions the "wash out" effect as well as the fast depletion of $H_2^+$ by the probe could be supressed and the predicted double-peak structure could be seen. Further details of our experimental setup are given in the Methods.

The process of ionization of $H_2$ by the pump-probe field is illustrated in figure 1(a). The final Coulomb explosion (CE) produces a pair of energetic protons, with kinetic energy release

(KER) given by the instantaneous positon of the NWP. That is, $KER = \frac{1}{R} + E_0$, where $E_0$ is the initial kinetic energy obtained in dissociation process [3] and R is inter-nuclear separation at the moment of second ionization. By changing the time delay, we can control the inter-nuclear separation at which ionization of the molecular ion takes place.

The measured KER spectra together with energy-integrated yields for the EI channel as a function of delay are shown in figure 2. The channel with KER around 10 eV comes from the sequential double ionization of $H_2$ by the pump pulse, and the channel with low KER of ~0.7 eV comes from the bond softening dissociation of $H_2^+$, followed by single ionization of the atomic H fragment. These channels are produced by the pump pulse alone and show no dependence on pump-probe delay. The channel with KER in the (2-6) eV range shows strong delay dependence and is attributed to ionization of $H_2^+$ by the probe pulse. The vanishing EI yield around zero delay proves that our laser pulse is sufficiently short and the pre- and post-pulses do not have enough intensity to induce enhanced ionization, so that the background signal of the EI yield produced by either pump or probe pulse alone is supressed. The EI signal becomes visible when the delay is increased to about 7 fs. From then on, two branches in the time-dependent KER spectra develop. The stronger branch has a decreasing central KER with increasing delay that originates from the dissociating NWP produced by the pump pulse. The weaker branch has a delay independent KER centred at around 4.5 eV. The energy-integrated yield of the EI channel presents two clear peaks at delays around 15 fs and 23 fs reflecting the R-dependent ionization probability for the dissociating molecular ion (figure 2(c)).

This time-dependent double-peak structure in the yield, in conjunction with the time-dependent KER identifying the channel, presents sufficient proof of the double-peak structure in R-dependent ionization probability. In order to relate the time delays to the inter-nuclear separations and to confirm our assignment we also performed a numerical simulation (see Methods b for details). The results of that simulation are presented in figure 3. The simulated delay-dependent KER spectra for enhanced ionization channel have the same structure as observed in the experiment (figure 3 (a)): a strong branch with delay dependent KER and a weak branch with delay-independent KER are present in the simulation as well. However, our model fails to quantitatively reproduce relative heights of the two peaks, most likely due to its reduced dimensionality with the electron motion constrained to the laser polarization direction. There are some theoretical indications that off-axis electron trajectories contribute significantly to the second peak [11], implying that a quantitative theoretical description will require a full-dimensional treatment of electrons.

It is known from theory and experiments that EI is very anisotropic, occurring only in molecules oriented within a narrow angle of laser polarization direction. Therefore, we would not expect to see any maximum in the time-dependent ionization yield for molecules aligned normal to laser polarization. To test this prediction we repeated our experiment, with the polarization axis of the probe pulse rotated by 90 degrees. Since the dissociating $H_2^+$ are preferentially created along the polarization axis of the pump pulse, this configuration probes molecules which are oriented perpendicular to the electric field. Fig 2(c) shows the results of the cross-polarized pump-probe experiment. The energy-integrated double-ionization yield increases monotonically with delay and remains constant after 15 fs while KER shows similar decrease with time as in the parallel-polarization case. The yield dependence is explained purely by decrease in ionization potential in dissociating molecular ions. The EI mechanism is not operational in perpendicular molecules. No peaks or double-peaks in ionization probability are expected and this is confirmed by our measurement.

The simulated dynamics of the dissociating NWP triggered by the pump pulse is shown in figure 4(a). We note that a dissociating NWP is propagating with a constant velocity of 0.35 a.u./ fs, with slight spatial expansion. In figure 4(a), the time-dependent expectation value for the inter-nuclear distance $<R_0(\tau)>$ is shown by the white dashed line. We also converted the experimental energy-dependent yield $Y(KER, \tau)$ into R-dependent yield $Y(R, \tau)$ using the relationship $KER = \frac{1}{R} + E_0$ where $E_0$ = 0.7 eV is the measured KER of the bond softening channel (fig. 4(b)). The same dashed white line as in fig. 4(a) is overlayed over the converted experimental spectra showing that the conversion truthfully reflects time evolution of NWPs. To extract the R-dependent ionization yield, we integrate the measured $Y(KER, \tau)$ over all delays. As shown in figure 4(d), a clear double peak structure is observed in the time-integrated KER spectrum, where the first peak is around 4.9 eV and the second peak is around 3 eV. By using the same conversion function of $KER = \frac{1}{R} + 0.7\ eV$, we can directly obtain inter-nuclear separations corresponding to the two peaks. The first peak is around 6.5 a.u., and the second peak is around 12 a.u., which agrees well with the fixed-nuclei model calculation (figure 4 (b), see Method c for details of the simulation).

In conclusion, we have presented the first experimental confirmation of the existence of the double-peak structure in R-dependent ionization rate of $H_2^+$, which has long been predicted by theory. We achieve this confirmation by performing a pump-probe experiment with a few-cycle probe pulse being 10 times weaker than the pump pulse. We determine that the ionization probability for $H_2^+$ is maximized at inter-nuclear separations of 6.5 a.u. and 12 a.u. in good agreement with theoretical predictions. Our numerical simulations agree well with the measurements and confirm our interpretation. Our results also confirm the validity and usefulness of the fixed-nuclei approximation for understanding ultrafast molecular dynamics. The well-resolved double peak structure also suggests that high degree of precision and control could be achieved in experiments on light diatomic molecules, opening new avenues in exploration of a variety of ultrafast processes and phenomena.

**Methods:**

a. **Experimental scheme:**

The schematic diagram of our experiment is shown in figure 1(c). A linear-polarized 6 fs 750 nm pulse is divided into a strong pump pulse and a weak probe pulse by a Mach-Zehnder interferometer, where the power of both pulses is individually controlled by apertures. The intensity of pump pulse ($6(\pm2)\times10^{14} W/cm^2$) and probe pulse ($6(\pm2)\times10^{13} W/cm^2$) at the laser focus is calibrated by making *in situ* measurements of the momentum of Ne$^+$ ions produced by circularly polarized pulses [15]. Our previous experiment [16] has shown that such laser parameters of the pump pulse can support strong radiative coupling between $\sigma_g$ and $\sigma_u$ states of $H_2^+$, which is essential for EI. The delay between pump and probe pulse is scanned from -5 fs to 50 fs with a step of 0.67 fs. The pulses were tightly focused by a silver-coated concave mirror ($f = 75\ mm$) installed inside a reaction microscope (REMI), onto a supersonic gas jet of hydrogen molecules. The polarization axis of the pump pulse is set to be parallel to the time-of-flight axis of REMI, and a half-wave plate is used to control the polarization of the probe pulse to be either parallel or perpendicular to polarization of the pump pulse. All the ions were detected by a time- and position-sensitive detector (Roendtek), and the three-dimensional momentum vectors were determined. We used a momentum conservation coincidence filter to select proton pairs from the same hydrogen molecule.

b. **Numerical simulation:**

We numerically solved the time-dependent Schrödinger equation (atomic units are used throughout unless indicated otherwise) [17]

$$i\frac{\partial}{\partial t}\Psi(z,R;t) = \left[\frac{p_R^2}{2\mu} + \frac{1}{2}(p_z + A(t))^2 + V(z,R) + \frac{1}{R}\right]\Psi(z,R;t)$$

where μ is the reduced nuclear mass, $p_R$ and $p_z$ are nuclei and electronic momentum operators, $A(t)$ is the laser vector potential, and $A(t) = -\int_0^t dt' E(t')$. The form of the electric field $E(t)$ will be presented later. To reproduce the molecular potential energy curves in such a reduced dimensionality model, we wrote the Coulomb potential in the form [18]:

$$V(z,R) = -\sum_{s=\pm1}\frac{1}{1/\alpha(R) - \alpha(R)/5 + \sqrt{(z+sR/2)^2 + (\alpha(R)/5)^2}}$$

where $\alpha(R)$ is the R-dependent soft-core-function. The simulation box extends from -3000 to 3000 a.u. along x dimension and from 0 to 40 a.u. along R dimension, and the spatial and time steps are $\Delta z = 0.3\ a.u., \Delta R = 0.02\ a.u., \Delta t = 0.2\ a.u.$. At the end of propagation t=t$_f$, we smoothly filtered out the wave function within the radius $\sqrt{R^2+z^2} = 60$ a.u. and Fourier transformed the remaining part into the momentum representation $\widetilde{\Psi}(p_R,z;t_f)$.

The proton energy spectrum is W(E) $\propto |\widetilde{\Psi}(p_R, z; t_f)|^2 / p_R$. We have checked that the simulation box is big enough that no wavepacket reached its boundaries during simulations, and the nuclear momentum and ionization probabilities converged. We assumed that the single ionization of $H_2$ induced by the pump pulse could be described by the Franck-Condon approximation. Hence, we launched the coherent superposition of first ten lowest molecular vibrational states of $H_2^+$, weighted by the Franck-Condon factors in the middle of the pump pulse. We obtained these ten states by propagating the wave function in imaginary time. The coherent nuclear wave packet moves to the outer turning point, and is dissociated by the tail of the pump pulse or the pedestal, followed by the ionization triggered by the time-delayed probe pulse. According to the experimental conditions, the pump pulse generally sits on the pedestal with long duration but a much lower intensity. In the simulations, we used the combined laser fields as following:

$$E = E_{pedestal} \sin(\omega t) + E_{probe} \sin[\omega(t - \Delta t)] f_{probe}$$

where

$$f_{probe} = \cos^2[\pi(t - \Delta t)/(5T)], \quad -2.5T < t - \Delta t < 2.5T$$

where T is the optical period, $E_{pedesta} = 0.01\ a.u.\ (3 \times 10^{12} W/cm^2)$, $E_{probe} = 0.0413\ a.u.\ (6 \times 10^{13} W/cm^2)$. The dissociating $H_2^+$ was produced by the pedestal when the inter-nuclear distance reached the critical value (about 4.7 a.u.) for bond softening.

c. *Fixed-nuclei model* simulation:

We numerically solved the time-dependent Schrödinger equation in cylindrical coordinates by taking the inter-nuclear distance R as a parameter,

$$i\frac{\partial}{\partial t}\Psi(\rho, z; t) = \left[\frac{p_\rho^2}{2} + \frac{1}{2}(p_z + A(t))^2 - \sum_{s=\pm 1}\frac{1}{\sqrt{(z + sR/2)^2 + \rho^2}}\right]\Psi(\rho, z; t).$$

The laser field is identical to the probe pulse in the Methods b, and the simulation algorithms are same as before. The simulation box is big enough that no wave packet reaches the numerical boundaries. We keep propagating the nuclear wave packet after the laser field is finished until all observations are converged. The ionization probability is counted by integrating the probability distributed in the area $\sqrt{\rho^2 + z^2} > 60$ a.u. at the end of the propagation. Compared to the moving nuclei simulation, here the description for the electron is complete considering the rotational symmetry.

Figure 1.

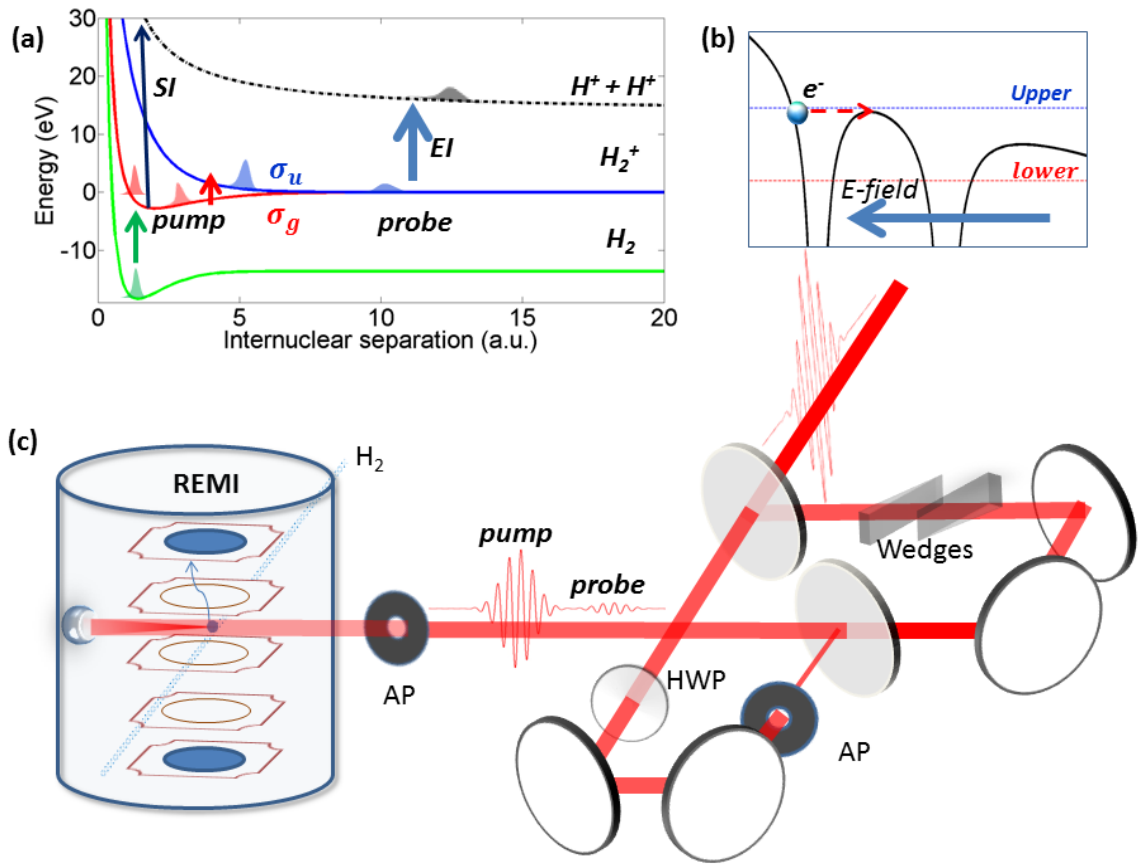

Figure 1. (a) Illustration of the enhanced ionization pathway when the hydrogen molecule interacts with an intense pump-probe laser field. Green arrow represents the tunnel ionisation of $H_2$ by the pump field, which launches a coherent Franck-Condon wavepacket from the ground state of $H_2$ (green line) onto the $\sigma_g$ curve of $H_2^+$ (red line). At the tail of the pump pulse, this wavepacket may be further excited to $\sigma_u$ curve of $H_2^+$ (blue line) by absorbing one photon from pump field (red arrow), producing a dissociating NWP. When the NWP reaches the critical region for enhanced ionization, the probe pulse projects (blue arrow) the NWP onto the Coulomb repulsive curve of $H_2^{2+}$. (b) Fundamental physical mechanism of EI. When $H_2^+$ with its double-well potential (black line) is exposed to a static electric field (blue arrow) the electron localized in the upper potential well will experience potential barrier suppressed by the Coulomb field of the second proton and it can much more easily tunnel out (red dashed arrow) than an atomic electron. (c) Experimental setup. REMI - reaction microscope; AP - adjustable aperture; HWP - half-wave plate.

Figure 2.

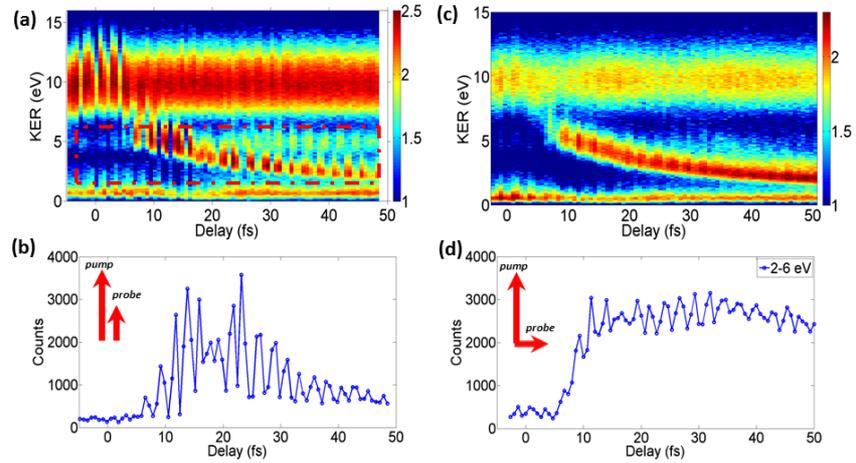

Figure 2. The measured delay-dependent KER spectra when pump and probe polarization axes are parallel (a) and perpendicular (c) to each other. The probe intensity for perpendicular polarization is approximately five times higher than for parallel one, as less intense pulses produced no significant EI signal in the cross-polarized configuration. Energy-integrated delay dependent ionization yields with parallel (b) and cross (d) polarized pump and probe. Note the double-peak structure for parallel polarization.

Figure 3

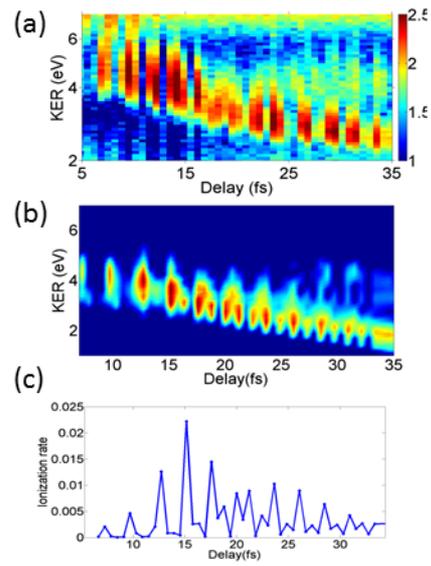

Figure 3. (a) Measured delay-dependent KER spectrum of the EI channel (same as in figure 2(a)) compared with the simulated delay-dependent KER spectrum (b). (c) The corresponding simulated delay-dependent energy-integrated ionization rate (see Methods for details). A periodic modulation of proton yield at a period of the driving laser (~2.5 fs) is also visible in the spectra. As confirmed by our numerical simulation, such modulation comes from the interference between the tail of the pump pulse and the probe pulse.

Figure 4

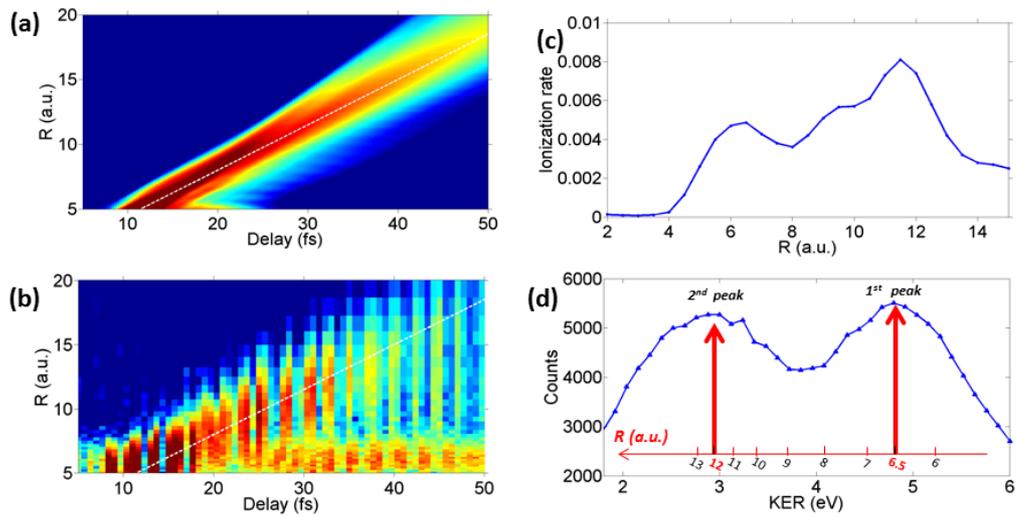

Figure 4. (a) Simulated time evolution of dissociating NWP density. The white dashed line represents the position of the NWP peak density as a function of delay. (b) Measured proton yield as a function of R and delay. The white dashed line is the same as shown in (a). (c) Simulated R-dependent ionization rate using fixed-nuclei model. (d) Measured KER spectrum integrated over all delays. The R axes in (b) and (d) are converted from KER by using $KER = \frac{1}{R} + 0.7 \ eV$.

**Acknowledgements**

This work was supported by an Australian Research Council (ARC) Discovery Project (DP110101894) and by the ARC Centre for Coherent X-Ray Science (CE0561787). H.X. was supported by an ARC Discovery Early Career Researcher Award (DE130101628). F.H thanks the financial support from NSFC (Grant No.11104180, 11175120, 11121504, 11322438). D.K. was supported by an ARC Future Fellowship (FT110100513).

Author contributions

H. X. and I. L. conceived of and designed the experiment. H.X performed the experiment and analysed the experimental data. F. H. performed the theoretical calculations. All authors contributed to discussions and preparation of the manuscript.